# Picosecond Ultrasonic Measurements Using an Optical Cavity

Y. Li, Q. Miao, A.V. Nurmikko and H.J. Maris

Department of Physics and Division of Engineering, Brown University, Providence, Rhode Island 02912, USA

A detailed analysis of the use of an optical cavity to enhance picosecond ultrasonic signals is presented. The optical cavity is formed between a distributed Bragg reflector (DBR) and the metal thin film samples to be studied. Experimental results for Al and Cu films show enhancement of acoustic signals by up to two orders of magnitude and are in good agreement with calculated results based on a previously established model. This technique provides an efficient method for detecting sound in materials with small piezo-optic coefficients and makes it possible to determine the actual pulse shape of the returning acoustic echoes.

# I. INTRODUCTION

Picosecond ultrasonics[1] has become a standard technique that can be used to perform a wide range of ultrasonic experiments on thin films and more complex nanostructures. A pump light pulse of duration typically around 1 picosecond is absorbed at the free surface of a metal or semiconducting film. This sets up a thermal stress which relaxes and sends a strain pulse into the film. This strain pulse travels to the far side of the film and is partially reflected there. When the reflected part of the pulse returns to the free surface of the film, it results in a small change $\Delta r_S$ in the optical reflectivity $r_S$ of the sample. This change is measured by means of a time-delayed probe light pulse. The change occurs because the strain pulse causes a change in the optical "constants" $n$ and $\kappa$ of the film. The initial experiments[1] measured the change in the intensity of the reflected probe light, i.e., the measured quantity was the change $\Delta R$ in the intensity reflection coefficient given by

$$\Delta R = |r_S + \Delta r_S|^2 - |r_S|^2. \quad (1)$$

It is also possible to detect the returning strain pulse through a measurement of the change in the phase of the reflected probe light.[2,3,4,5] This change arises because when the sound wave returns there is a change $\Delta \phi_S$ in the phase of the reflection coefficient and a change $\Delta z$ in the position of the upper surface of the film. Other possibilities are to measure the change in the polarization of the reflected probe light or a change in the transmitted light (only possible if the film is partially transmitting).

There are two significant challenges in making these measurements. The first is simply that the strain amplitude is usually quite small and so the change in the optical properties at the film surface are also small. The value of $\Delta R$ in most experiments performed to date is in the range $10^{-5}$ to $10^{-6}$. To measure a change of this magnitude requires the use of signal averaging and lock-in techniques.

The second problem is that for some combinations of film material and light wavelength the value of $\Delta R$ happens to have a very small value. For example, for light of wavelength 800 nm (the most common wavelength of commercial compact ultrafast lasers) the change in the reflectivity of copper with strain is close to zero.[6] Thus, either extensive signal averaging has to be performed or the phase change of the reflected probe light has to be measured instead of the intensity. A number of different methods have been developed to measure this change in the phase; these methods involve different types of interferometer. One difficulty with standard interferometers is that the distance of optical components from the sample has to be controlled very precisely. However, it has been shown that this difficulty can be overcome, for example, by using a modified Sagnac interferometer.[3]

In this paper we describe another method that can be used to improve the signal measured in picosecond ultrasonic experiments and discuss its range of applicability. In this method we place a reflector immediately above the sample surface so that an optical cavity is formed in the space between the reflector and the sample. The spacing of this cavity is chosen so that it is at close to resonance. A change in the reflection coefficient of the sample will change the reflectivity $\Delta R$ of the probe light pulse. By making the cavity have as high a Q as possible and by choosing the cavity spacing so that the probe wavelength is at an optimal point on the cavity resonance, the value of $\Delta R$ can be maximized. Our discussion will primarily be in the context of ultrasonic measurements but the same techniques could be used for a wide variety of ultrafast optical pump-and-probe experiments.

An optical cavity has been used to enhance sensitivity in many experiments within different disciplines such as Raman spectroscopy[7,8] and biosensing down to single molecules.[9] In several previous experiments optical cavities have been used to enhance ultrasonic signals.[10] However, in those experiments samples were fabricated with an optical cavity being an integral part of the sample structure. Here our goal is to investigate the use of a versatile <u>external</u> optical cavity formed by a high reflectance element which is brought into close proximity for measurements to a reflecting test sample. This arrangement enables detailed picosecond ultrasonic characterization for measurements on a wide range of test materials, including copper, for potential use in metrology for the semiconductor industry.

## II. DESIGN CONSIDERATIONS

We consider the basic equations governing detection with an optical cavity (Fig. 1). The cavity consists of the sample film (reflectivity $r_S$) and the reflector. We choose the free surface of the sample to be at $z = w$ and the sample to lie in the region $z > w$. We take the reflector to lie entirely in the region $z < 0$, so the space between $z = 0$ and $z = w$ is vacuum. The probe light has a wavelength in vacuum of $\lambda_0$ and an angle of incidence of $\theta$. Then the overall reflectivity of the reflector plus sample is

$$r = r_{RP} + \frac{t_R^2 r_S \exp(2ik_z w)}{1 - r_{RN} r_S \exp(2ik_z w)}, \quad (2)$$

where $k_z$ is the z-component of the wave vector of the probe light inside the cavity, i.e., $2\pi \cos\theta / \lambda_0$, $r_{RP}$ is the reflection coefficient of light incident on the reflector going in the positive z-direction, $r_{RN}$ is for light going in the negative z-direction, $t_R$ is the transmission coefficient, and $r_S$ is the reflection coefficient of the sample. The intensity reflection coefficient is then $R = |r|^2$. If the unperturbed values of $r_{RP}$, $r_{RN}$, $t_R$, $r_S$ and $w$ are known, it is straightforward to calculate the change $\Delta R_1$ in $R$ resulting from a change in the magnitude of $r_S$, the change $\Delta R_2$ from a change in the phase $\phi_S$ of $r_S$, and the change $\Delta R_3$ due to a displacement $\Delta w$ of the surface of the sample. Clearly, the effect of a change $\Delta \phi_S$ in $\phi_S$ produces the same result as a change in $w$ of

$$\Delta w = \frac{\Delta \phi_S}{2k_z}. \quad (3)$$

We will consider the reflector to be composed entirely of some number of dielectric films with no light absorption.

It follows from time-reversal invariance that regardless of the details of the structure of the reflector, the reflection and transmission coefficients can always be written in the form

$$r_{RP} = -\tanh\mu \, \exp i(\varepsilon - \zeta) \quad r_{RN} = \tanh\mu \, \exp i(\varepsilon + \zeta) \quad t_R = \exp(i\varepsilon)/\cosh\mu, \quad (4)$$

The value of the parameter $\zeta$ is affected by the position of the reflector, whereas $\varepsilon$ and $\mu$ are determined by its structure. Using Eq. 4 in Eq. 2 then gives

$$r = \exp[i(\varepsilon - \zeta)]\left[-|r_R| + \frac{|r_S|(1-|r_R|^2)\exp(i\alpha)}{1-|r_R||r_S|\exp(i\alpha)}\right], \quad (5)$$

where

$$\alpha = 2k_z w + \varepsilon + \zeta + \phi_S, \quad (6)$$

and $|r_R|$ is the magnitude of the reflection coefficient of the reflector. In the experiments to be described in this paper the values of $\varepsilon$, $\phi_S$ and $\zeta$ are not known (the choice of the reference plane position $z_R$ is not even specified). For this reason we have to work in terms of an effective cavity width $w_{eff} \equiv w + \frac{\varepsilon + \zeta + \phi_S}{2k_z}$. From Eq. 5

$$R = \frac{(|r_R|-|r_S|)^2 + 4|r_R||r_S|\sin^2(\alpha/2)}{(1-|r_R||r_S|)^2 + 4|r_R||r_S|\sin^2(\alpha/2)}. \quad (7)$$

The minimum value of the reflectivity is

$$R_{min} = \frac{(|r_R|-|r_S|)^2}{(1-|r_R||r_S|)^2}, \quad (8)$$

and always occurs when $\sin(\alpha/2) = 0$, i.e., when

$$w_{eff} = n\lambda_z/2, \quad (9)$$

where

$$\lambda_z = 2\pi/k_z = \lambda_0/\cos\theta. \quad (10)$$

In an experiment, the directly measured quantity is the change in the intensity of the reflected probe light, i.e., a change proportional to $\Delta R$. However, because normally the main source of noise in an experiment arises from the random fluctuations in the intensity of the probe beam, the signal to noise ratio is determined by $\Delta R/R$.[11] For this reason we will focus on optimizing this parameter.

For given properties of the sample, i.e., given values of $|r_S|$ and $\Delta r_S$, it is interesting to consider how to choose $|r_R|$ in order to make $\Delta R/R$ as large as possible. As an example, we consider a sample that has an intensity reflectivity $R_S = |r_S|^2$ of 0.85. The reflectivity of the cavity for different values of $R_R = |r_R|^2$ is shown in Fig. 2. One can see that as $R_R$ is increased the width of the cavity resonance decreases. Note also that, as can be seen from Eq. 7, the reflection at resonance is zero if $R_R = R_S$. The maximum field intensity in the cavity is when $R_R = R_S$ and $\alpha$ equal to an even multiple of $\pi$. We can divide $\Delta R$ into two components. The first component $\Delta R_1$ arises from a transient change $\Delta |r_s|$ in the magnitude of $|r_S|$, and the second $\Delta R_2$ the change $\Delta \alpha$ in $\alpha$.

In Fig. 3 we show $\frac{1}{R}\frac{dR}{d|r_S|}$ for the same set of parameters as used in Fig. 2. The magnitude of $\frac{1}{R}\frac{dR}{d|r_S|}$ is largest when the cavity is at resonance, i.e., when $\alpha = 0$ (or an even integer times $\pi$). At resonance the sensitivity is

$$\frac{1}{R}\frac{dR}{d|r_S|} = \frac{2\ (1-|r_R|^2)}{(|r_S|-|r_R|)(1-|r_S||r_R|)}. \tag{11}$$

This gives an infinite sensitivity when $|r_R| = |r_S|$; the reflectivity itself is zero when this condition is satisfied. Note that the quantity $\frac{1}{R}\frac{dR}{d|r_S|}$ is a discontinuous function of $\alpha$ and $R_R$ when both $\alpha$ and $R_R - R_S$ are zero.

To determine the extent to which the introduction of the optical cavity increases the change in the reflectivity, we use Eq.11 to obtain the result

$$\frac{dR}{R} = \frac{\sqrt{R_S}\ (1-R_R)}{\left(\sqrt{R_S}-\sqrt{R_R}\right)\left(1-\sqrt{R_R R_S}\right)}\frac{dR_S}{R_S} \equiv G\frac{dR_S}{R_S}, \tag{12}$$

so the gain $G$ is the ratio of the fractional change in reflectivity $dR/R$ with the cavity to the fractional change in reflectivity $dR_S/R_S$ of the sample itself. As an example, in Fig. 4 we show the magnitude of the gain as a function of $R_R$ for a sample with reflectivity $R_S = 0.85$. To achieve a very large gain it is necessary to have $R_R$ close to $R_S$ which is undesirable since it gives a very small overall reflection coefficient. However, for example, as a compromise one can choose $R_R = 0.94$ which gives a reflection coefficient of 0.2 and a gain of magnitude 11. Note that when $R_R = R_S$ the gain remains finite as $\alpha \to 0$, even though $R$ goes to zero.

In Fig. 5 we show plots of $\frac{1}{R}\frac{dR}{d\alpha}$, the parameter that determines how the reflectivity of the cavity changes in response to changes in the phase $\phi_S$ of $r_S$ or due to a displacement $\Delta w$ of the surface of the sample. Again, the sensitivity of the cavity increases with increasing $|r_R|$, but, as can be seen from Fig. 1, the width of the resonance decreases and so the range of $\alpha$ in which the sensitivity is high decreases.

These calculations do not include two important factors that limit the gain that can be obtained through the use of the cavity. These are the finite wavelength spread of the laser used for the probe light and the spread in the angle of incidence of the light. Consider, for example, a probe light pulse that has a time profile of intensity which is a Gaussian with a full width at half maximum of $\tau_{FWHM}$. Then because of the spread in wavelengths the effective reflection coefficient of the pulse will be

$$R_{eff} = \frac{\int \exp\left[-(\lambda-\lambda_0)^2/(\Delta\lambda^2)\right] R(\lambda) d\lambda}{\int \exp\left[-(\lambda-\lambda_0)^2/(\Delta\lambda^2)\right] d\lambda} \tag{13}$$

where $\lambda_0$ is the center wavelength of the pulse, $R(\lambda)$ is the reflectivity for wavelength $\lambda$, and

$$\Delta\lambda = \frac{\lambda_0^2 \sqrt{\ln 2}}{\pi c \tau_{FWHM}}. \tag{14}$$

For $\lambda_0 = 800$ nm and $\tau_{FWHM} = 100$ fs, $\Delta\lambda = 5.7$ nm. Clearly, if $\Delta\lambda$ is comparable to the width of the resonance of the cavity the maximum gain that can be obtained using the cavity will be decreased. Note too that the effect of a spread in wavelength is minimum for the lowest resonant mode of the cavity ($w_{eff} = \lambda/2$) and becomes progressively larger for the higher order modes.

To consider the effect of the spread of angles of the probe light note that as far as the cavity resonance is concerned the relevant parameter is the normal component of the light wave vector, i.e., $2\pi\cos\theta/\lambda$. Thus, when $\theta$ has a range $\Delta\theta$ this is equivalent to a wavelength range of $\Delta\lambda = \lambda_0 \tan\theta \Delta\theta$. For a probe beam with $\theta = \pi/4$ and with $\Delta\theta$ of 0.1 radians this gives a very large value of $\Delta\lambda$, i.e., 80 nm for $\lambda_0 = 800$ nm. However, the effect of a spread in angles can easily be minimized by using probe light at normal incidence. In this case when the angle of the probe beam extends from zero to $\Delta\theta/2$, the equivalent spread in wavelength is

$$\Delta\lambda = \lambda_0 \, (\Delta\theta)^2 / 8 \qquad (15)$$

Thus, for $\Delta\theta = 0.1$, $\Delta\lambda$ is only 1 nm.

The use of the cavity also increases the part of the energy of the pump light that is absorbed by the sample. This increase is large for highly reflecting samples such as copper.

### III. EXPERIMENT

Measurements were made using a conventional pump and probe experimental setup. The laser was a mode locked Ti:Sapphire laser (Spectra Physics Mai Tai VF-N1-06) that produces 800 nm pulses at a repetition rate of 80 MHz. The pulse width was 65 fs. The laser output was split into pump and probe beams using the combination of a $\lambda/2$ wave plate and a polarization sensitive beam splitter. The pump beam was modulated at a frequency $f_{mod} = 1.7$ MHz by an electro-optic modulator. The detected probe light was down-converted to 20 kHz using a mixer and then fed into a lock-in amplifier. Both light beams were passed through laser line filters[12] to narrow the wavelength bandwidth of the light from 12.5 nm to 3.2 nm. The advantage of this is discussed below. The pump and probe beams were focused by two objective lenses to a - spot on the sample. The probe beam was at normal incidence and the pump beam was at approximately 0.07 radians from normal. The spot size was measured by moving a knife edge across the focus. The intensity of both beams at a distance $r$ from the center of the spot was reasonably well described by $I_0 \exp(-r^2/\xi^2)$ where $I_0$ is the intensity at the center and $\xi = 11$ μm. As a reflector, we used a commercial dielectric Bragg reflector (DBR) fabricated to have a reflectivity of 0.84 for 800 nm light at normal incidence. The reflectivity was a maximum at 800 nm and varied by less than 0.04 over the range between 750 nm and 850 nm.

Two metal samples were studied. An aluminum film of thickness 190 nm was prepared by rf sputtering in a pressure of $10^{-9}$ torr. The substrate was sapphire. A copper film of thickness 180 nm was prepared under the same conditions with a silicon substrate.

The roughness of the surface of both films was less that 10 nm rms as measured by a white light interferometer.

The DBR was placed directly on top of the metal film, and was at a small angle to the film. By measuring the spectrum of white light reflected from the DBR/sample, we could determine the spacing between the DBR and the sample surface by tracking the Fabry-Perot resonances. This spacing was determined at a number of points and was assumed to vary linearly between these points. The angle between the film and the DBR was determined by the change of measured cavity spacing versus the translational displacement along the direction of the wedge. It was necessary to be sure that the angle between the DBR and the metal film was less than $10^{-4}$ rads in order that the cavity spacing did not vary by a significant amount over the area of the spot onto which the pump and probe beams were focused.

The effect of the laser line filter on the reflectivity of probe light from the optical cavity with the copper sample is shown in Fig. 6. One can see that when the laser line filter is used to narrow the band width of the probe light the depth of each minimum in the reflectivity is nearly the same, whereas without the filter the depth of successive minima decreases with increasing cavity spacing. This is to be expected from Eqs. 13 and 14 and the related discussion given earlier. Similar measurements were made for the other film. From the measured minimum reflectivity when the line filter was used, we can find from Eq. 8 a value for the reflectivity of the metal film. The results were 0.793 and 0.955 for the Al and Cu films, respectively.

To achieve a maximum acoustic signal one would like the pump light to have as low a reflection coefficient as possible and for the probe to have a reflection coefficient that varies as rapidly as possible. For this reason there is an advantage to having the pump and probe at slightly different angles, as in the present experiment. However a large angle difference is not desirable because it would shift the pump resonance too far away from the probe resonance and undo the effect of resonant absorption of the pump light. It would also be possible to use two laser line filters to divide the spectrum of the laser output into pump and probe pulses of slightly different wavelength.

In Fig. 7, we show the reflectivity of the pump and the probe light from the cavity with the Al sample. The cavity spacing is in the vicinity of the 5$^{th}$ resonance. One can see that the minimum reflectivity occurs at different spacing for the pump and the probe. This is because the probe light is at normal incidence while the pump light is at an angle $\theta$ of approximately 0.07 radians. This should shift the resonance by

$$\Delta w \approx w \frac{\theta^2}{2} = 5 \text{ nm}, \tag{16}$$

and this is in reasonable agreement with the data in Fig. 7.

## IV. RESULTS AND DISCUSSION

In Fig. 8a we show the results of pump and probe measurements on the Al film without the cavity. One can see a series of acoustic echoes with time spacing 57 ps superposed on a smoothly varying background signal. The background arises because of the transient heating of the structure by the pump pulse. The sign of successive echoes changes; this is because the acoustic strain changes sign when the sound is reflected at the free surface of the Al, but is unchanged when reflection occurs at the Al/sapphire interface. The power in the modulated pump beam was approximately given by $[30 + 19\cos(2\pi f_{\text{mod}} t)]$ mW. Since the reflectivity of the Al film was 0.793, the average pulse energy absorbed in the film was 0.078 nJ, and the amplitude of the modulation in the absorbed pulse energy was $\Delta Q = 0.049$ nJ.

In Fig. 8b we show results for the same Al film when the cavity is used with the same incident pump power. The spacing of the cavity has been chosen to maximize the magnitude of $\Delta R/R$ for the first acoustic echo. In Fig. 9, we show results for $\Delta R/R$ for 9 different cavity spacings with the background due to transient heating subtracted.[13] Curve number 6 corresponds to the data shown in Fig. 8b. The largest value of $|\Delta R|/R$ is $4 \times 10^{-4}$ which is bigger than the value of $\Delta R/R$ measured without the cavity by a factor of 170. Part of this increase arises simply because more pump power is absorbed in the Al film when the cavity is used. The reflectivity of the cavity for the pump light pulse for curve number 6 is 0.09. The DBR is a pure reflector and does not absorb light. Hence the fraction of the incident pump light that is absorbed is increased from 0.207 when the cavity is not used, to 0.91, and so the amplitude of the modulation in the absorbed pulse energy is $\Delta Q = 0.341$ nJ. If instead the same amount of pump energy was absorbed in the Al film, the cavity would give a signal larger by a factor of 45.

The precise piezo-optic coefficients for thin films of aluminum are likely to depend on the deposition techniques and environmental effects, such as surface oxidation. Jiles and Staines[14] have measured these coefficients for an aluminum film as a function of wavelength. However, their results for the derivatives vary very rapidly with wavelength in the vicinity of 800 nm and, in addition, the values of $n$ and $\kappa$ that they found differ considerably from the values $n = 1.17$ and $\kappa = 4.15$ found on our sample.[15] Presumably, this is due to differences in the method for film preparation. Consequently, we have adopted the following method. From the results of ref. 1, it is straightforward to show that the change in the sample reflection coefficient due to a strain pulse is given by

$$\Delta |r_S| = \int_0^\infty h(z) \eta(z,t) \, dz, \tag{17}$$

$$\Delta \phi_S = \int_0^\infty g(z) \eta(z,t) \, dz, \tag{18}$$

where $\eta$ is the $zz$ component of the strain tensor,

$$h(z) = \frac{2 k_0^3}{|k - k_0||k + k_0|} \mathrm{Re}\left\{ r_S^* \frac{\partial \varepsilon}{\partial \eta} \exp[i(2 n k_0 z + \alpha)] \right\} \exp(-z/\zeta), \tag{19}$$

$$g(z) = \frac{2 k_0^3}{|k - k_0|^2} \mathrm{Im}\left( r_S^* \frac{\partial \varepsilon}{\partial \eta} \exp[i(2 n k_0 z + \alpha)] \right) \exp(-z/\zeta) - 2 k_0, \tag{20}$$

and $\varepsilon = (n + i\kappa)^2$, $k$ is the wave number of the light inside the sample ($= k_0 \sqrt{\varepsilon}$), and $\zeta = \lambda_0 / 4\pi\kappa$. From Eq. 17, the fractional change in the intensity reflection from the film (no cavity) can be calculated as

$$\frac{\Delta R_{film}}{R_{film}} = \frac{2 \Delta |r_S|}{|r_S|}. \tag{21}$$

The fractional change in reflection when the cavity is used is

$$\frac{\Delta R_{cavity}}{R_{cavity}} = \frac{1}{R_{cavity}} \frac{d R_{cavity}}{d |r_S|} \Delta |r_S| + \frac{1}{R_{cavity}} \frac{d R_{cavity}}{d \phi_S} \Delta \phi_S. \tag{22}$$

We make a fit to the first acoustic echo that is centered around 57 ps. Let the strain associated with the returning sound pulse that gives this echo be $\eta_1(z + vt)$, where $t$ is measured from the time at which the center of the pulse reaches the surface of the film. This left going pulse is reflected at the free surface and so the total strain will be $\eta_1(z+vt) - \eta_1(-z+vt)$. We describe the function $\eta_1$ by its value at $N$ values of its argument. These values, together with the values of the strain derivatives $\partial n / \partial \eta$ and $\partial \kappa / \partial \eta$, were adjusted to give the best possible fit to the data taken with and without the cavity. Data set #6 (see Fig. 9) was used for the cavity. In making this fit we used the measured reflectivity of the cavity (Fig. 7) to obtain

$$\frac{1}{R_{cavity}} \frac{dR_{cavity}}{d|r_S|} = 0.28 \qquad \frac{1}{R_{cavity}} \frac{dR_{cavity}}{d\phi_S} = 12.7 \qquad (23)$$

The values obtained for the piezo-optic constants are $\partial n / \partial \eta = -0.8$ and $\partial \kappa / \partial \eta = 4.5$. The results of this fit are shown in Figs. 10-12. The strain pulse as determined in this way is shown by the open circles in Fig. 10; this is for the Al film without the cavity, i.e., it is for an absorbed pulse energy $\Delta Q$ of 0.049 nJ. When the cavity is used the shape of the pulse should be the same but the amplitude will be larger because a greater fraction of the pump energy is absorbed. The open circles in Figs. 11 and 12 show the measured first echo $\Delta R(t)/R$ for the Al film with and without the cavity, respectively. The solid curves are the results of calculations of $\Delta R(t)/R$ based on the strain shape as given in Fig. 10, the use of Eqs. 17-23, and the values $\partial n / \partial \eta = -0.8$ and $\partial \kappa / \partial \eta = 4.5$. It can be seen that a very good fit to the experimental reflectivity data is obtained.

We now compare the result just obtained for the strain pulse with theory. The energy deposited by the pump pulse per unit volume of the sample is

$$\frac{\Delta Q}{A\zeta} \exp(-z/\zeta), \qquad (24)$$

where $A$ is the area illuminated by the pump and probe. In metals of high conductivity, the energy transferred to the electrons from an absorbed light pulse can rapidly diffuse a distance significantly larger than $\zeta$ before the energy is transferred to the thermal phonon bath. As a rough approximation, one can take the energy profile to still be exponential but with an effective absorption length $\zeta'$ that is larger than $\zeta$. Then following from the results in ref. 1, the pump light sets up a stress in the metal film which has a $zz$-component

$$\sigma_{zz} = -\frac{3B\beta \Delta Q}{CA\zeta'} \exp(-z/\zeta') \qquad (25)$$

where $B$ is the bulk modulus, and $\beta$ is the coefficient of linear thermal expansion. This initial stress results in a strain pulse propagating into the sample. At a time $t$ after the pump light pulse has been applied and before the pulse has reached the far side of the film, the $\eta_{33}$ component of the strain tensor is

$$\eta_{33} = \eta_0 \exp(-z/\zeta') - \frac{1}{2}\eta_0 \exp[-(z-vt)/\zeta'] - \frac{1}{2}\eta_0 \exp[-(z+vt)/\zeta'] \quad z > vt$$
$$= \eta_0 \exp(-z/\zeta') + \frac{1}{2}\eta_0 \exp[(z-vt)/\zeta'] - \frac{1}{2}\eta_0 \exp[-(z+vt)/\zeta'] \quad z < vt \quad (26)$$

where

$$\eta_0 = \frac{\beta \, \Delta Q}{CA\zeta'} \frac{1+\sigma}{1-\sigma}, \quad (27)$$

$v$ is the sound velocity, and $\sigma$ is Poisson's ratio. When the sound has gone across the metal film and been reflected at the interface to the substrate, there is a returning echo. The form of this echo is[16]

$$\eta_{33} = -\frac{1}{2}\eta_1 \exp[(z-z_1)/\zeta'] \quad z < z_1$$
$$= \frac{1}{2}\eta_1 \exp[-(z-z_1)/\zeta'] \quad z > z_1 \quad (28)$$

where the center of the pulse is at $z_1 = 2d - vt$, $d$ is the thickness of the metal film, and

$$\eta_1 = r_{AC}\eta_0, \quad (29)$$

with $r_{AC}$ the acoustic reflection coefficient at the interface between the metal film and the substrate. For Al, $\beta = 2.2 \times 10^{-5}$ K$^{-1}$, $C$=2.4 J cm$^{-3}$K$^{-1}$, and $\sigma = 0.35$, and the acoustic reflection coefficient at the interface between Al and sapphire is 0.44. It is straightforward to show that the effective area when the pump and probe beams have Gaussian profiles is $A = 2\pi\xi^2$. Using these parameters and a value of $\zeta'$ of 50 nm, we obtain $\eta_1 = 1.08 \times 10^{-5}$ and the results for the first acoustic echo that are shown in Fig. 10. This is in very reasonable agreement with the experimental result, considering that $\zeta'$ is the only adjustable parameter involved, and that no allowance has been made for the broadening of the pulse due to attenuation.

Figure 13 shows pump and probe data for the copper sample taken with the cavity; the background contribution has been subtracted in the way described below. For the copper sample, no acoustic echoes can be seen when the cavity is not used. This is to be expected since Gerhardt[6] has measured the piezo-optic coefficients of copper and found that at 800 nm the coefficients were zero to within the measurement accuracy. Consequently, we assume that the entire signal arises from the surface displacement, i.e., we assume that $\Delta |r_S|$ and $\Delta \phi_S$ can both be neglected.

Because of the high diffusion coefficient for electrons in copper, when the pump light pulse is absorbed a stress is set up that extends throughout the copper film. If we suppose that this decreases as $\exp(-z/\zeta')$ from the front of the film, the stress is

$$\sigma_{zz} = -\frac{3B\beta \Delta Q}{CA\zeta'} \frac{\exp(-z/\zeta')}{1-\exp(-W/\zeta')}, \quad (30)$$

where $W$ is the thickness of the film. This reduces to Eq. 25 when $W \gg \zeta'$ and the factor in the denominator is included so that the integral of the stress over the film thickness has the correct value. This stress gives rise to strain pulses propagating in the positive and negative $z$-direction. It is straightforward to show that the surface displacement is

$$u_S = -\frac{\beta \Delta Q}{CA} \frac{1+\sigma}{1-\sigma}\left[1 - r_{ac}^n \frac{\exp(-|vt-2nW|/\zeta')-\exp(-W/\zeta')}{1-\exp(-W/\zeta')}\right] \quad (31)$$

when $t$ lies in the interval between $(2n-1)W/v$ and $(2n+1)W/v$, with $n$ an integer 0, 1, 2... One can see from this formula that the surface displacement reaches a maximum negative value whenever the time is an odd integer times the time for sound to travel through the film. For Cu, $\beta = 1.7\times10^{-5}$ K$^{-1}$, $C$=3.45 J cm$^{-3}$K$^{-1}$, $v = 4.73\times10^{5}$ cm s$^{-1}$, and $\sigma = 0.345$. The reflection coefficient at the interface to the substrate is 0.366 and, based on the sound velocity and the measured round trip time, the film thickness is 180 nm. For the cavity spacing that gives the largest signal ($w=1606$ nm), the amplitude of the modulation in the absorbed pulse energy is $\Delta Q = 0.14$ nJ, and the sensitivity of the cavity to changes in width is $(1/R)(dR/dw) = -0.12$ nm$^{-1}$. Using these values together with Eq. 31 and $\zeta'$ values of 100 and 150 nm gives the theoretical curves shown in Fig. 13. To compare the calculation with experiment it is necessary to subtract from the experimental data the smoothly varying background term that arises from the change in the optical reflectivity of the sample due to the change in temperature (thermoreflectance). For the aluminum sample it is straightforward to do this because the sound signal appears as rather sharp echoes (see Fig. 8). For copper, on the other hand, the sound echoes are broad and so it is not so easy to distinguish them from the background. The result of approximating the background as a constant plus a decaying exponential[17] gives the results shown in Fig. 13. One can see that the echoes seen experimentally are smaller and broader than predicted by the calculation. We cannot tell whether this is the result of attenuation (this is large in copper because of the elastic anisotropy of the grains making up the film), or the approximation for the initial stress (Eq. 30).

## V. SUMMARY

In this paper we have investigated the use of an optical Fabry-Perot cavity to enhance the signals that are detected in picosecond ultrasonic experiments. We have discussed the considerations involved in the design of the cavity and have presented results of measurements made using this technique. The use of a cavity has several important advantages over the standard technique where the optoacoustic generation and detection relies on a single thin film "transceiver". The signal can be enhanced by a significant factor - up to two orders of magnitude; the maximum possible enhancement depending on the reflectivity and piezo-optic coefficients of the sample. Important to practical applications in testing and metrology, the method does not require that the laser used for the probe light have a wavelength at which the sample has large piezo-optic coefficients. For example, for copper (key material for interconnect wiring in the semiconductor industry) this is a significant advantage because this material has essentially no piezo-optic response at the standard wavelength of 800 nm for many commercial ultrafast lasers. The cavity technique should make it possible to perform measurements using low cost short pulse semiconductor or fiber lasers, including those developed for the optical telecommunication industry near at 1.5 microns. Finally, we note that, provided the displacement component makes the main contribution to the signal, it is now possible to determine the actual pulse shape of the returning acoustic echoes, thereby enriching the total amount of information acquired in psec ultrasonic experiments.

This work was supported by the Air Force Office of Scientific Research under contract FA FA9550-08-1-0340 through the Multidisciplinary University Research. We thank T. Grimsley and F. Yang for helpful discussions.

## FIGURE CAPTIONS

Fig. 1. Schematic diagram of the optical cavity, formed from an external reflector placed in proximity ($\sim \lambda$) and parallel to a reflecting test sample surface.

Fig. 2. The optical intensity reflectivity $R$ of the cavity as a function of the parameter $\alpha$ defined in Eq. 6. The intensity reflectivity of the sample is 0.85. The curves are labeled by the values of the intensity reflectivity of the reflector $|r_R|^2$.

Fig. 3. The change in the intensity reflectivity $R$ of the cavity due to a small change in the magnitude of the amplitude reflection coefficient $r_S$ of the sample as a function of the parameter $\alpha$. The intensity reflectivity of the sample is 0.85. The different curves are labeled by the values of the intensity reflection coefficient of the reflector $R_R$.

Fig. 4. The gain in sensitivity $G$ due to an optical cavity as a function of the intensity reflection coefficient $R_R$ of the reflector. The cavity is assumed to be at resonance. The dashed curve is the overall intensity reflectivity coefficient $R$ of the structure. The intensity reflectivity of the sample is 0.85.

Fig. 5. The change in the reflectivity of a cavity with respect to the change in the parameter $\alpha$ defined in Eq. 6. The different curves are labeled by the values of the intensity reflectivity of the reflector $R_R$.

Fig. 6. Measured reflectivity of the optical cavity with the copper film as a function of the cavity spacing. Circles are measurements using the laser line filter to narrow the spectrum of the probe light, and the squares are without using the filter.

Fig. 7. Measured reflectivity of the pump (circles) and probe light (squares) for the optical cavity with the aluminum film.

Fig. 8. a) The change $\Delta R(t)$ in the reflectivity of the Al film as a function of time after the application of the pump light pulse in a "standard" psec ultrasonic experiment. b) Results obtained for the same film when the cavity is used. Note especially the change in vertical scale.

Fig. 9. Measured values of $\Delta R(t)/R$ for the aluminum film as a function of the probe delay time when the optical cavity is used. The different curves are data for a sequence of measurements with increasing cavity spacing. A smoothly varying background contribution has been subtracted from each data set.

Fig. 10. The open circles show the strain of the first returning acoustic pulse in the aluminum sample without using the cavity. The shape is determined through the analysis based on Eqs. 17-23. The solid curve is the result of the calculation based on Eqs. 24-29.

Fig. 11. The open squares show the shape of the first acoustic echo $\Delta R(t)/R$ in the aluminum film when measured without using the cavity. The solid curve is the result of the fit based on Eqs. 17-23.

Fig. 12. The open squares show the shape of the first acoustic echo $\Delta R(t)/R$ in the aluminum film when measured using the cavity. The solid curve is the result of the fit based on Eqs. 17-23. The dashed and dotted curves show the contributions from the displacement of the film surface and the piezo-optic effect, respectively.

Fig. 13. $\Delta R(t)/R$ for the copper film when measured using the optical cavity. The solid line shows the experimental data after background subtraction of a constant plus a decaying exponential. The dotted and dashed curves show the value of $\Delta R(t)/R$ calculated from Eq. 31 using values of $\zeta'$ of 100 and 150 nm, respectively.

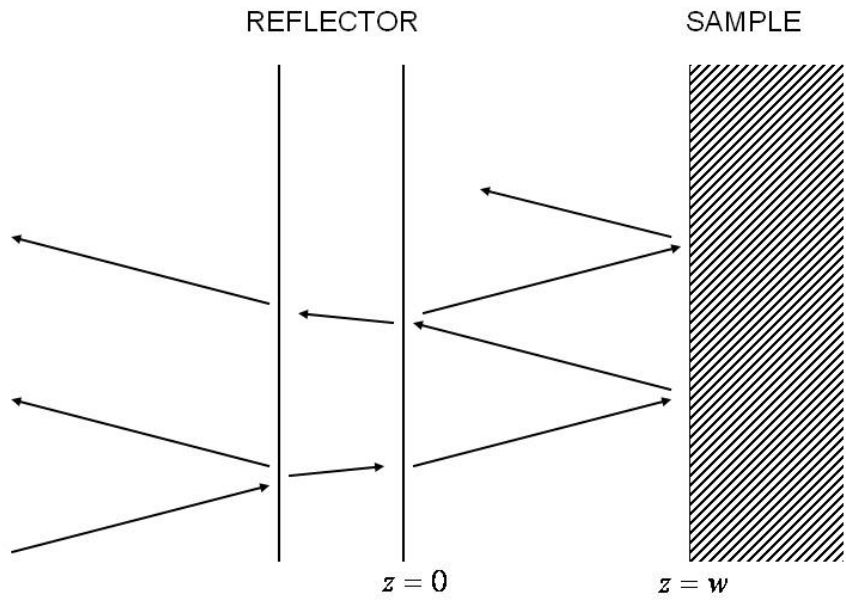

Fig. 1

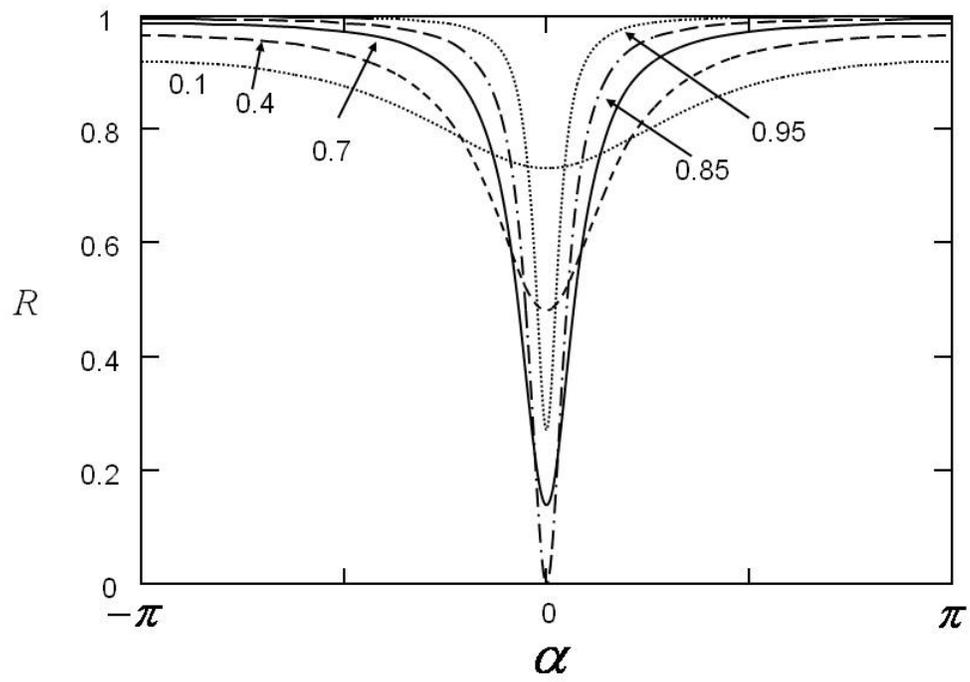

Fig. 2

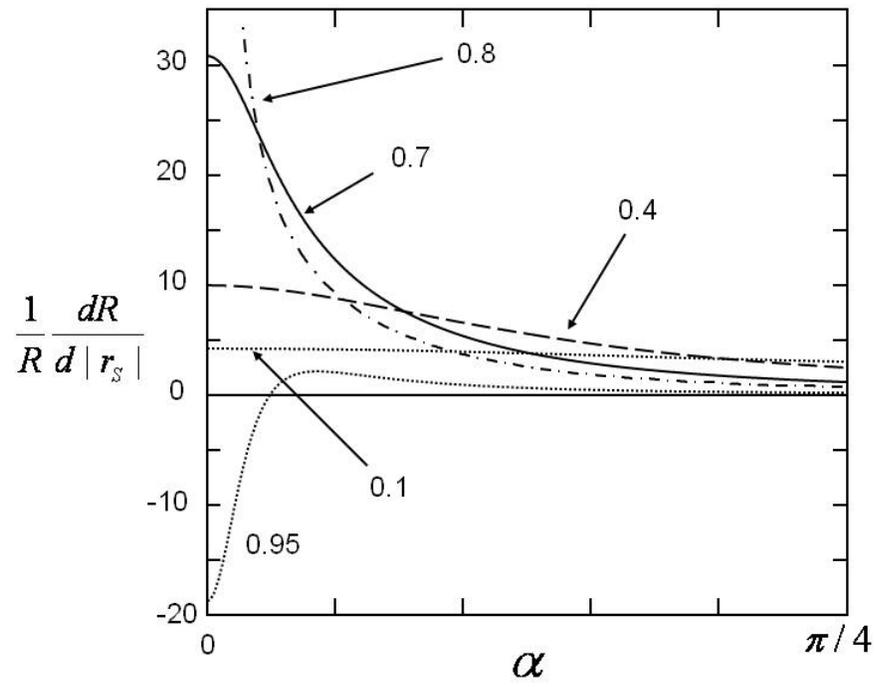

Fig. 3

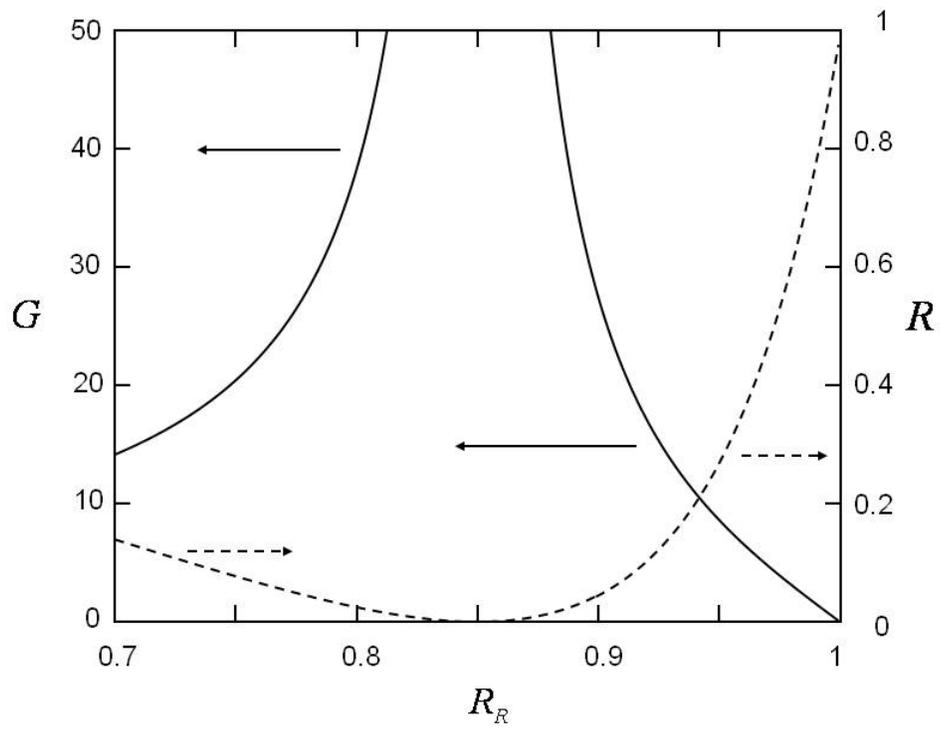

Fig. 4

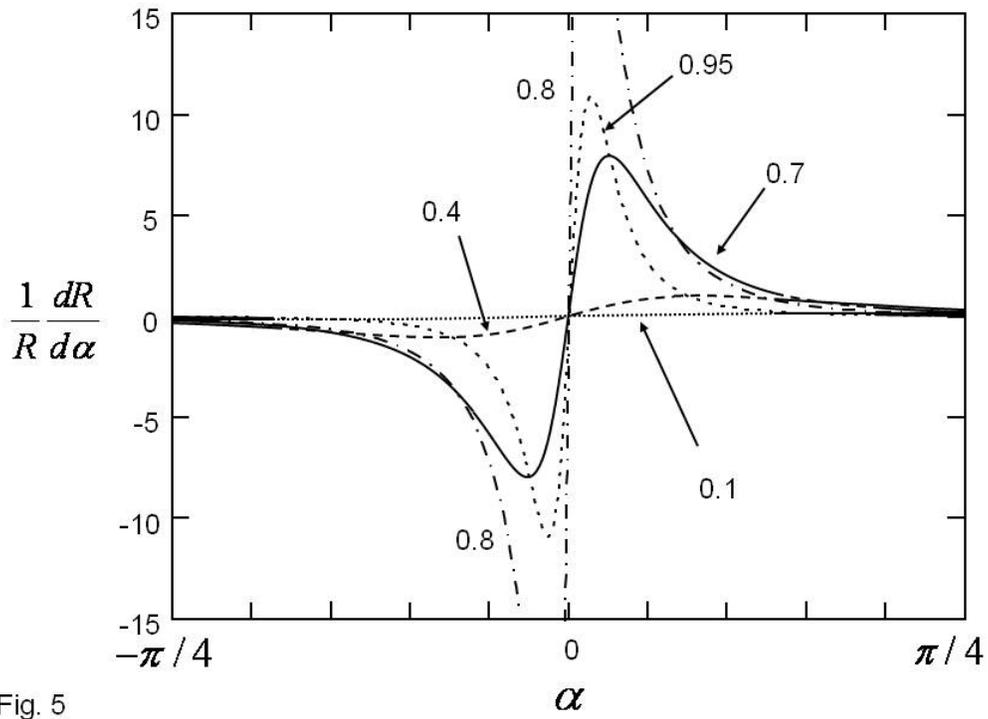

Fig. 5

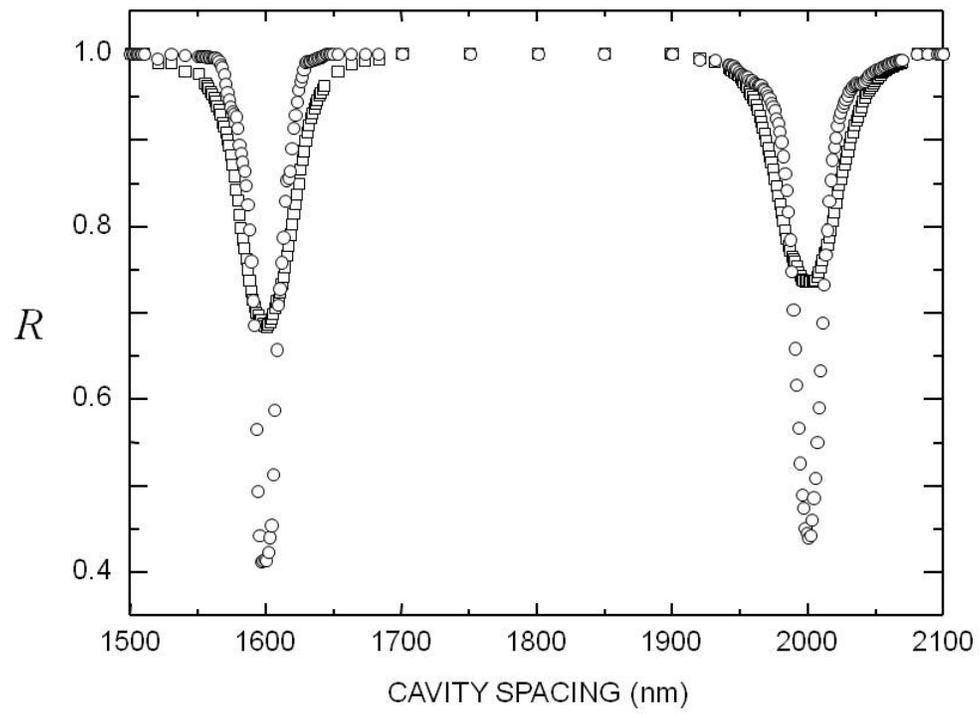

Fig. 6

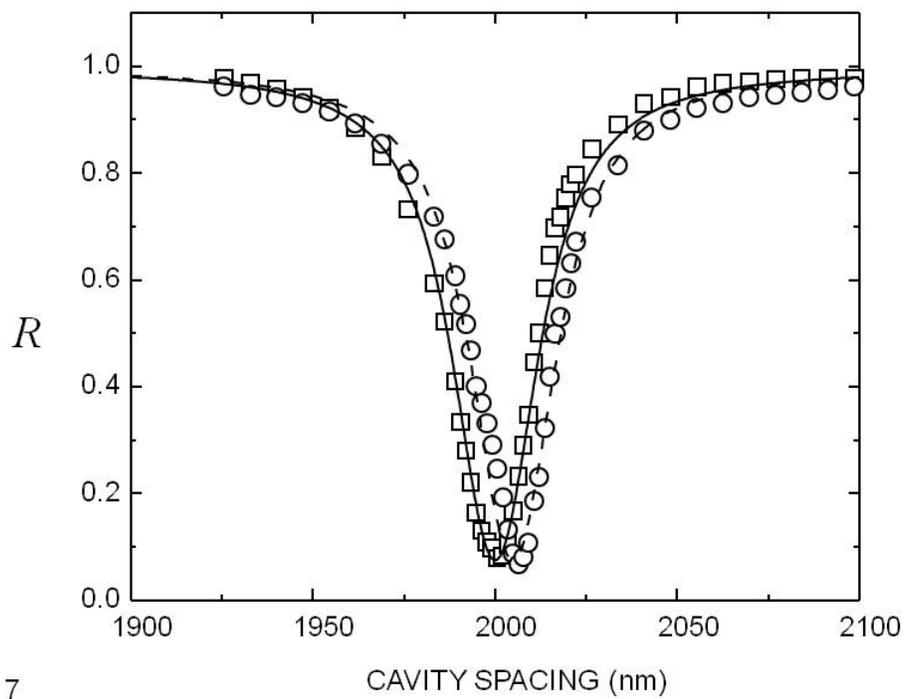

Fig. 7

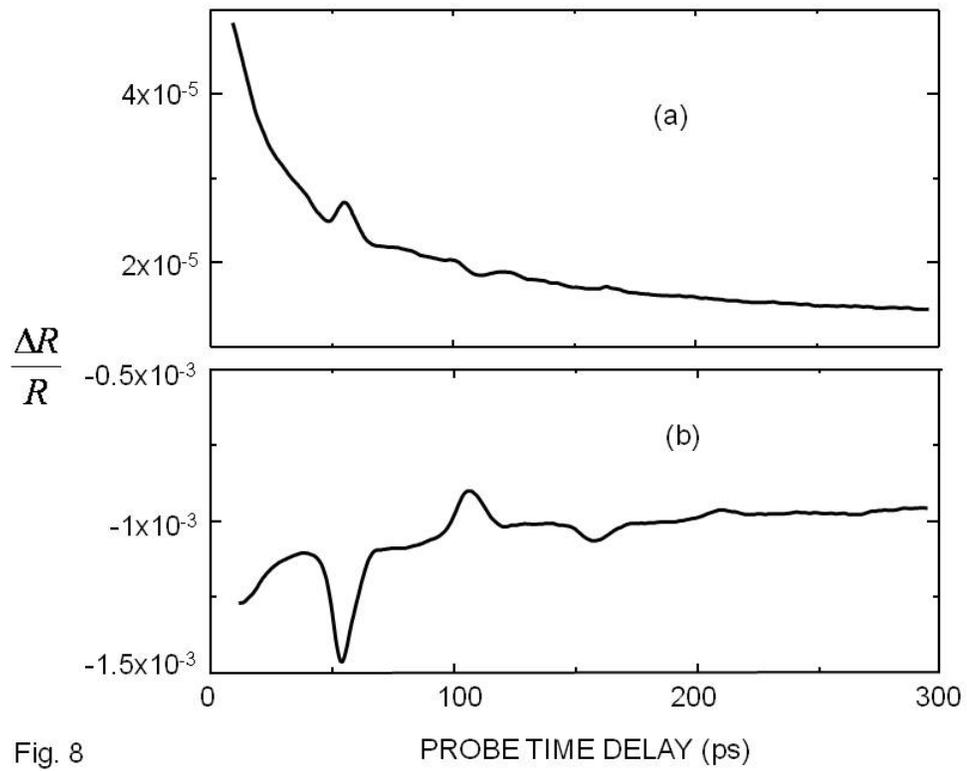

Fig. 8

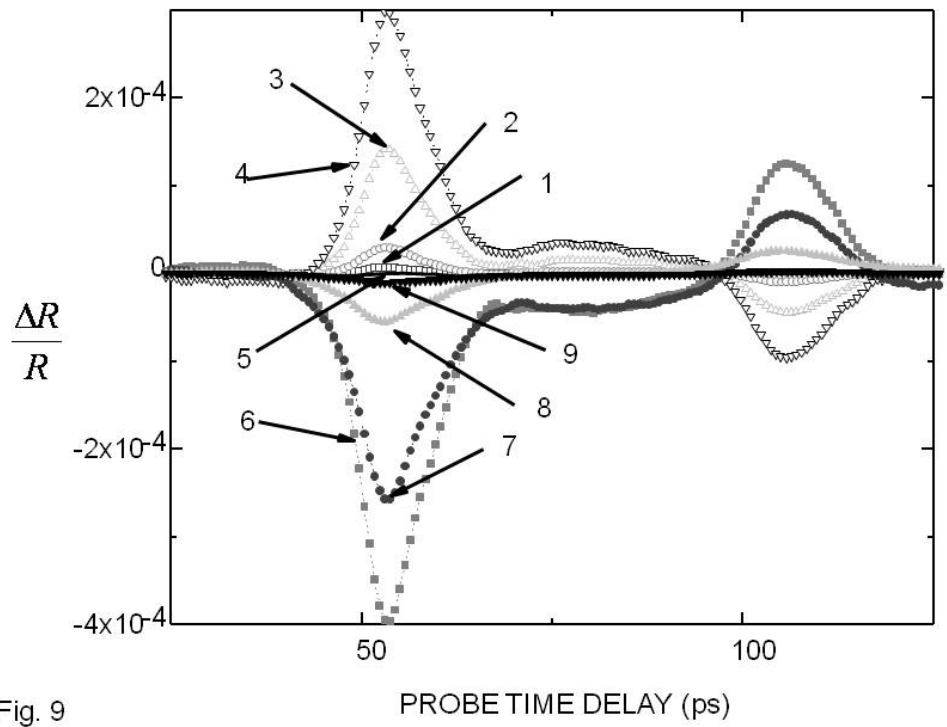

Fig. 9

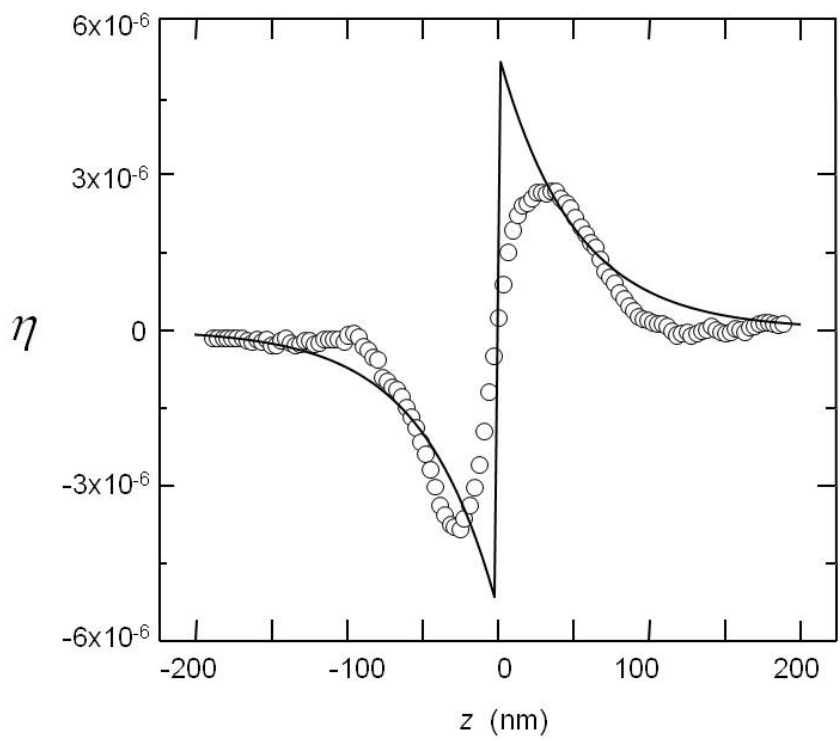

Fig. 10

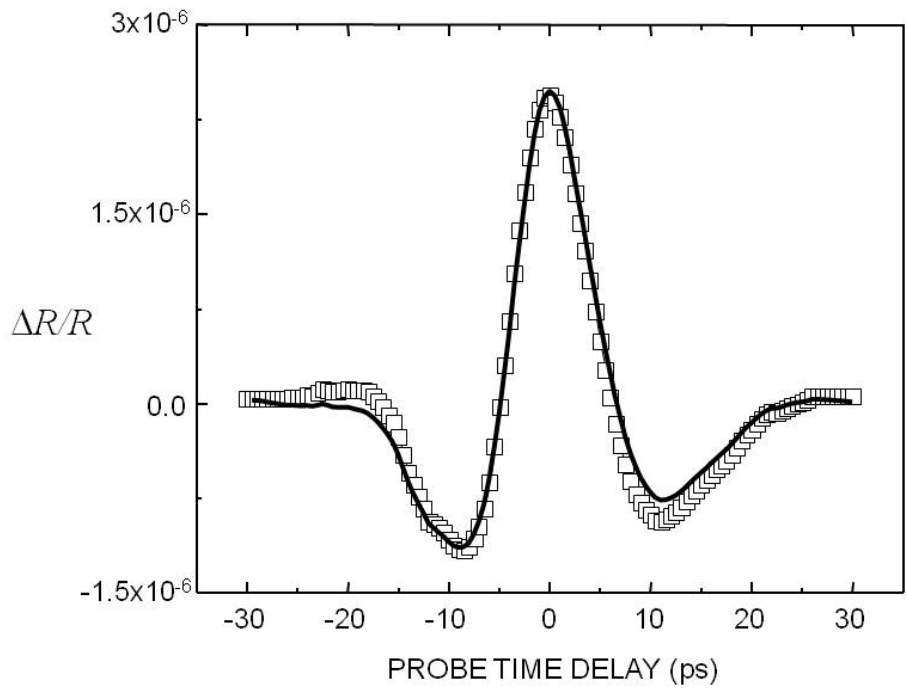

Fig. 11

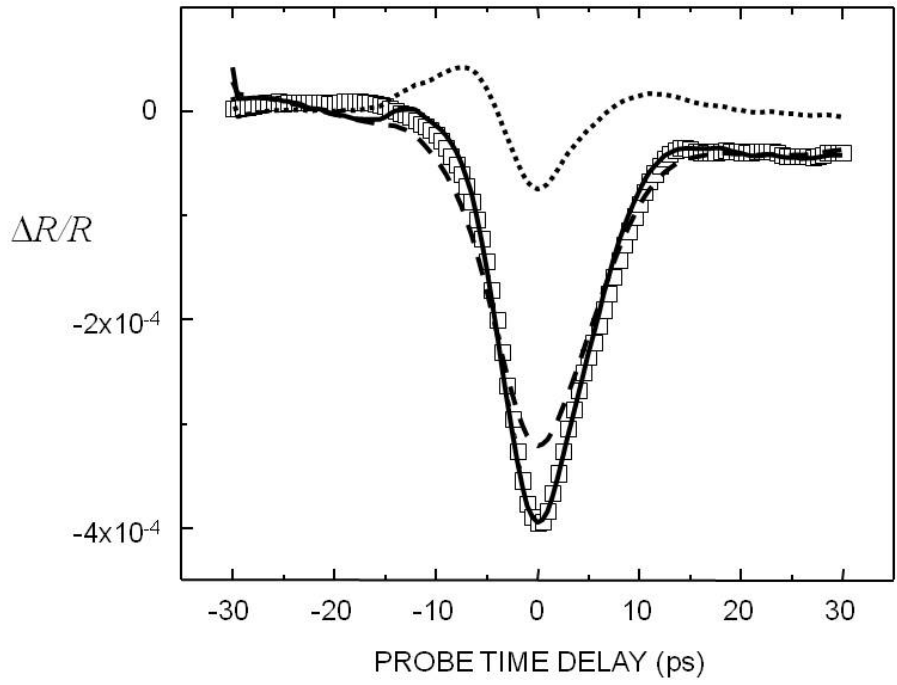

Fig. 12

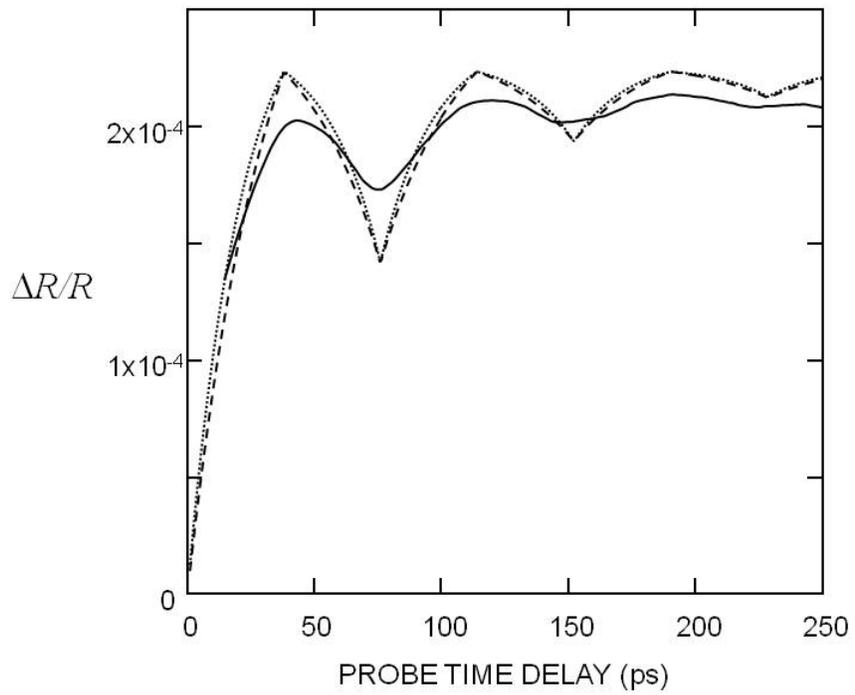

Fig. 13